\documentclass[%
reprint,
superscriptaddress,
amsmath,amssymb,
aps,prl,
]{revtex4-2}
\usepackage{amsmath,amssymb}
\usepackage{graphicx}
\usepackage{dcolumn}
\usepackage{bm}
\usepackage{multirow}
\usepackage{float}
\usepackage{hhline}
\usepackage{color}
\usepackage{afterpage}
\usepackage{tabularx}
\usepackage{comment}
\usepackage{float}
\usepackage{mathtools}
\usepackage[utf8]{inputenc}
\usepackage[T1]{fontenc}
\usepackage{mathptmx}
\usepackage[normalem]{ulem}
\usepackage{physics}
\usepackage[caption=false]{subfig}

\newcolumntype{C}[1]{>{\hsize=#1\hsize\centering\arraybackslash}X}%
\newcolumntype{L}[1]{>{\hsize=#1\hsize\raggedright\arraybackslash}X}%

\graphicspath{{Figures/}}

\DeclareMathAlphabet\mathcal{OMS}{cmsy}{m}{n}

\begin{document}

\preprint{AIP/123-QED}

\title{Polaron-Polariton Assisted Thermally Activated Superradiance}

\author{Yi-Ting Chuang}
\email{yiting.chuang.research@gmail.com}
\affiliation{Institute of Atomic and Molecular Sciences, Academia Sinica, Taipei 10617, Taiwan}
\affiliation{Department of Chemistry, National Taiwan University, Taipei 10617, Taiwan}
\author{Liang-Yan Hsu}
\email{lyhsu@gate.sinica.edu.tw}
\affiliation{Institute of Atomic and Molecular Sciences, Academia Sinica, Taipei 10617, Taiwan}
\affiliation{Department of Chemistry, National Taiwan University, Taipei 10617, Taiwan}
\affiliation{Physics Division, National Center for Theoretical Sciences, Taipei 10617, Taiwan}

\begin{abstract}
We predict an anomalous thermally activated superradiance in molecular aggregates within polaritonic environments. In contrast to free space, the collective emission is enhanced when either the exciton-phonon coupling or the temperature increases. This counterintuitive phenomenon is captured by a microscopic theory that combines macroscopic quantum electrodynamics with a modified polaron quantum master equation approach, revealing a nontrivial interplay among excitons, phonons, and polaritons.
\end{abstract}

\maketitle

\textit{Introduction}.--- 
Molecular aggregates have attracted significant attention due to their unique optical properties and potential applications in light-emitting devices \cite{Saikin2013, Hestand2017, Hestand2018, Zhang2020, Zhao2020}. One of their most striking features is superradiance—the enhancement of collective spontaneous emission rate beyond that of an individual monomer \cite{Dicke1954}. Pioneering studies of molecular superradiance trace back to Kasha’s seminal works on excitons in aggregates, which revealed that Coulombic dipole-dipole interactions among monomers lead to distinct photophysical behaviors depending on the packing configuration of the monomers \cite{Kasha1958, Kasha1959, Kasha1963, Kasha1965}. Subsequent studies have extended Kasha’s framework to include exciton-phonon interactions \cite{Spano1991, Spano2010, Schröter2015, Hestand2018} and disorder \cite{Fidder1991, Spano2005}, both of which typically diminish superradiance. \textcolor{black}{In addition, superradiance in molecular aggregates is generally suppressed as the temperature increases \cite{DeBoer1990, Fidder1990, Fidder1991b, Fidder1995, Kamalov1996, Monshouwer1997, Scheblykin2000, Palacios2002, Lecuiller2002, Manna2019, Manna2020}. To the best of our knowledge, thermally activated superradiance has rarely been observed and is typically anticipated only in molecular aggregates where the bright state lies energetically above dark states \cite{Yamagata2012, Spano2014}, as reported in H aggregates \cite{Renger2009} and two-dimensional aggregates \cite{Eisfeld2017}. 
In contrast, this phenomenon is not expected in one-dimensional J aggregates, where such high-lying bright-state configurations are absent.}

Recently, coupling molecules to confined electromagnetic fields, such as cavity photons or polaritons (photons dressed by dielectric environments), has emerged as a novel approach for modifying molecular processes, including energy transfer \cite{Andrew2004, Coles2014, Zhong2017, Wu2018, Li2021} and chemical reactions \cite{Hutchison2012, Semenov2019, Thomas2019, Li2020, Wei2024, Sharma2024}. In particular, polariton-assisted superradiance in a J aggregate has been theoretically predicted to yield anomalous giant rate enhancement, exceeding that in free space by roughly an order of magnitude \cite{Chuang2024b}. However, the previous study has not accounted for the vibrational and thermal effects, which are known to suppress superradiance in molecular aggregates. \textcolor{black}{To overcome these limitations, we develop a general theory of single-photon superradiance that describes molecular aggregates coupled to both phonons and polaritons at finite temperatures. This theory combines macroscopic quantum electrodynamics (MQED) \cite{Gruner1996, Dung1998, Buhmann2012, Hsu2025}—a first-principles quantization scheme for electromagnetic fields in inhomogeneous, dispersive and absorbing media—with the polaron quantum master equation approach \cite{Jang2011, Nazir2016}.} We show that, in contrast to free space, superradiance in molecular aggregates coupled to polaritons can be enhanced by increasing either the exciton-phonon coupling or the temperature. Our results reveal an unconventional thermally activated superradiance mechanism in one-dimensional J aggregates and open new directions for controlling molecular emission via dielectric engineering.

\textit{Model Hamiltonian}.---
We consider molecular aggregates, composed of a set of $N$ identical electronic two-level systems (TLSs), interacting simultaneously with phonons (originate from intramolecular and local environmental vibrational modes) and polaritons \textcolor{black}{(photons dressed by the dielectric environment)}. Based on the multipolar coupling framework of MQED, the total Hamiltonian under the electric-dipole approximation can be expressed as 
\begin{align}
\nonumber
    \hat{\mathcal{H}} = & \sum_{\alpha} \hbar\omega_{\mathrm{M}} \hat{\sigma}^{+}_\alpha \hat{\sigma}^{-}_\alpha + \sum_\alpha\sum_q \hbar \omega_{\mathrm{V}, \alpha q} \hat{b}^\dagger_{\alpha q} \hat{b}_{\alpha q} \\
\nonumber
    & + \sum_\alpha \hat{\sigma}^{+}_\alpha \hat{\sigma}^{-}_\alpha \left\{ \sum_q \hbar \omega_{\mathrm{V},\alpha q} \left[ \sqrt{S_{\alpha q}} (\hat{b}^\dagger_{\alpha q} + \hat{b}_{\alpha q}) + S_{\alpha q} \right] \right\} \\
    & + \int \mathrm{d}\mathbf{r} \int_{0}^{\infty} \mathrm{d}\omega \, \hbar\omega \,\mathbf{\hat{f}}^\dagger(\bf{r},\omega)\cdot\mathbf{\hat{f}}(\bf{r},\omega) - \sum_\alpha \hat{\boldsymbol{\mu}}_{\alpha} \cdot \hat{\mathbf{F}}(\mathbf{r}_\alpha).
\label{Eq:H_tot}
\end{align}
The first term describes the TLSs constituting the aggregate, where $\omega_{\mathrm{M}}$ represents the transition frequency of each monomer, and $\hat{\sigma}^{+}_\alpha = \ket{\mathrm{e}_\alpha}\bra{\mathrm{g}_\alpha}$ ($\hat{\sigma}^{-}_\alpha = \ket{\mathrm{g}_\alpha}\bra{\mathrm{e}_\alpha}$) denotes the raising (lowering) operator for monomer $\alpha$, with $\ket{\mathrm{g}_\alpha}$ and $\ket{\mathrm{e}_\alpha}$ corresponding to the electronic ground and excited states of monomer $\alpha$, respectively. The second term describes the phonon (vibrational) modes, where $\omega_{\mathrm{V}, \alpha q}$ and $\hat{b}_{\alpha q}$ denote the frequency and the annihilation operator of the $q$-th phonon mode that couples to monomer $\alpha$. The third term describes the interactions between the monomers and the phonon modes, where $S_{\alpha q}$ denotes the Huang-Rhys factor that characterizes the coupling between monomer $\alpha$ and its phonon mode $q$. We consider that each monomer couples individually to a separate phonon bath and that both the phonon bath properties and exciton-phonon coupling strengths are identical across all monomers; therefore, the notation can be further simplified as $\omega_{\mathrm{V}, \alpha q} = \omega_{\mathrm{V}, q}$ and $S_{\alpha q} = S_{q}$, and the exciton-phonon coupling is determined by the phonon (vibrational) spectral density $J_\mathrm{V}(\omega) = \sum_q S_{q} \omega_{\mathrm{V}, q}^2 \delta(\omega - \omega_{\mathrm{V}, q})$. The fourth term describes the polaritons \textcolor{black}{(photons dressed by the dielectric environment)}, where $\mathbf{\hat{f}}(\mathbf{r},\omega)$ denotes the annihilation operator of the bosonic vector fields \textcolor{black}{(polariton fields)}. The fifth term describes the interactions between the monomers and polaritons, where $\hat{\boldsymbol{\mu}}_\alpha = \boldsymbol{\mu}_\alpha (\hat{\sigma}^{+}_{\alpha} + \hat{\sigma}^{-}_{\alpha})$ denotes the transition dipole operator of monomer $\alpha$, and $\boldsymbol{\mu}_\alpha$ represents its corresponding transition dipole moment. The field operator $\hat{\mathbf{F}}(\mathbf{r}_\alpha)$ is defined as $\hat{\mathbf{F}}(\mathbf{r}_\alpha) = \int \dd{\mathbf{r}} \int_0^\infty \dd{\omega}  \overline{\overline{\mathcal{G}}}(\mathbf{r}_\alpha,\mathbf{r},\omega) \cdot \hat{\mathbf{f}}(\mathbf{r},\omega) + \mathrm{H.c.}$, where the tensor $\overline{\overline{\mathcal{G}}}(\mathbf{r}_\alpha,\mathbf{r},\omega)$ is explicitly given by $\overline{\overline{\mathcal{G}}}(\mathbf{r}_\alpha,\mathbf{r},\omega) = i\sqrt{\frac{\hbar}{\pi\varepsilon_0}} \frac{\omega^2}{c^2} \sqrt{\mathrm{Im} \left[ \varepsilon_\mathrm{r}(\mathbf{r},\omega) \right]} \, \overline{\overline{\mathbf{G}}}(\mathbf{r}_\alpha,\mathbf{r},\omega)$. The symbols $\varepsilon_0$, $\varepsilon_\mathrm{r}(\mathbf{r},\omega)$, and $c$ denote the permittivity of free space, the relative permittivity, and the speed of light in free space, respectively; the tensor $\overline{\overline{\mathbf{G}}}(\mathbf{r},\mathbf{r'},\omega)$ denotes the dyadic Green's function that satisfies macroscopic Maxwell’s equations \textcolor{black}{(Maxwell's equations in matter)} $\left[ \frac{\omega^2}{c^2}\varepsilon_\mathrm{r}(\mathbf{r},\omega) - \nabla \times \nabla \times \right] \overline{\overline{\mathbf{G}}}(\mathbf{r},\mathbf{r'},\omega) = -\mathbf{\overline{\overline{I}}}_3 \delta(\mathbf{r}-\mathbf{r'})$, where $\mathbf{\overline{\overline{I}}}_3$ is the three-dimensional identity matrix. The information regarding the exciton-polariton coupling is fully encoded within the generalized polariton spectral density $\overline{\overline{\mathbf{J}}}_\mathrm{P}(\omega)$, with elements $J_{\mathrm{P}, \alpha \beta}(\omega) = \frac{\omega^2}{\pi \hbar \varepsilon_0 c^2} \boldsymbol{\mu}_\alpha \cdot \mathrm{Im} \overline{\overline{\mathbf{G}}}(\mathbf{r}_\alpha, \mathbf{r}_\beta,\omega) \cdot \boldsymbol{\mu}_\beta$.

\begin{figure}[!t]
    \centering
    \includegraphics[width=0.47\textwidth]{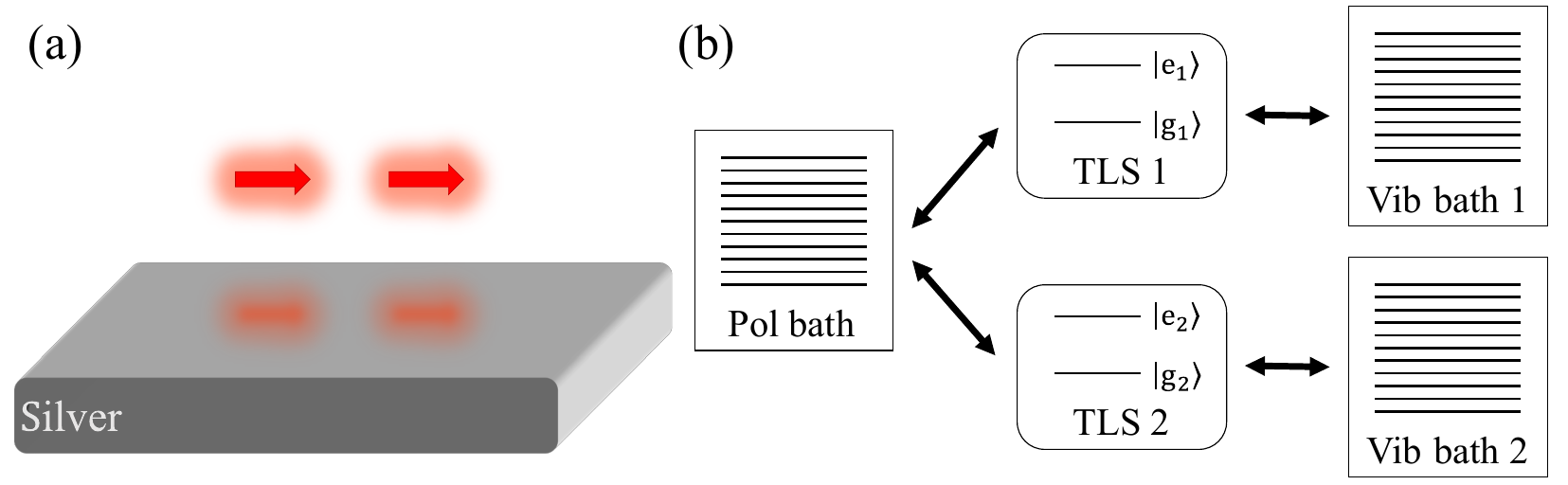} 
    \caption{(a) Schematic of a J aggregate composed of two monomers above a silver surface. \textcolor{black}{(b) Illustration of the physical model: each monomer consists of a two-level system (TLS) coupled to its phonon (Vib) bath, and all monomers are coupled to a global polariton bath.}}
    \label{Fig1}
\end{figure}

\textit{\textcolor{black}{Generalized theory of superradiance}}.---
Our goal is to explore the combined effect of exciton-phonon and exciton-polariton couplings on the collective decay rates of the aggregate (excited) states. Following previous studies \cite{Spano1989, Chuang2024b}, the aggregate states $\ket{\tilde{\mathrm{E}}_{k}}$ are defined as
$\ket{\tilde{\mathrm{E}}_k} = \sum_{\alpha} U_{\alpha k} \ket{\mathrm{E}_{\alpha}}$, where the local excited state of monomer $\alpha$ is expressed as $\ket{\mathrm{E}_{\alpha}} = \hat{\sigma}^{+}_\alpha \ket{\mathrm{G}}$, with $\ket{\mathrm{G}} = \prod_\alpha \ket{\mathrm{g}_\alpha}$. The coefficients $U_{\alpha k}$ correspond to elements of the unitary matrix $\overline{\overline{\mathbf{U}}}$ that diagonalizes the free-space dipole-dipole interaction matrix  $\overline{\overline{\mathbf{V}}}\vphantom{a}^0$, whose matrix elements are explicitly given by $V^0_{\alpha \beta} = -\frac{\omega_{\mathrm{M}}^2}{\varepsilon_0 c^2} \boldsymbol{\mu}_{\alpha} \cdot \mathrm{Re}\overline{\overline{\mathbf{G}}}\vphantom{a}^0(\mathbf{r}_\alpha,\mathbf{r}_\beta,\omega_{\mathrm{M}}) \cdot \boldsymbol{\mu}_{\beta}$ for $\alpha\neq\beta$ and $V^0_{\alpha \beta} = 0$ for $\alpha=\beta$, with $\overline{\overline{\mathbf{G}}}\vphantom{a}^0(\mathbf{r}_\alpha,\mathbf{r}_{\beta},\omega_{\mathrm{M}})$ denoting the free-space dyadic Green's function \cite{Chuang2024}. The state for which the coefficients $U_{\alpha k}$ exhibit the same sign for all $\alpha$ is further referred to as the superradiant or bright state, owing to its maximal collective decay rate. 

\begin{figure*}[!t]
    \centering
    \includegraphics[width=0.98\textwidth]{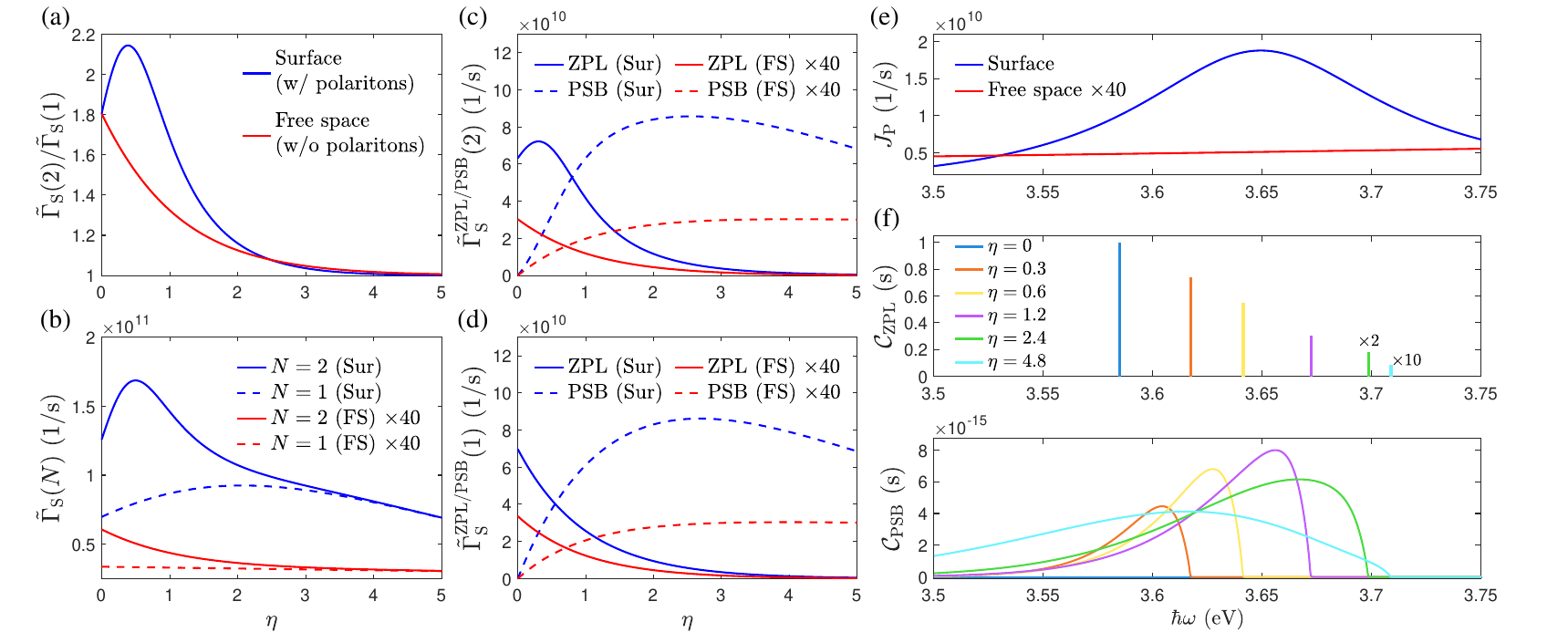} 
    \caption{\textcolor{black}{Interplay between exciton-phonon coupling and exciton-polariton coupling on superradiance at $T=0~\mathrm{K}$. (a) Exciton-phonon coupling strength $\eta$ dependence of the superradiance rate enhancement $\tilde{\Gamma}_\mathrm{S}(2)/\tilde{\Gamma}_\mathrm{S}(1)$ for a dimer above a silver surface (with polaritons) and in free space (without polaritons). (b) $\eta$ dependence of the rates $\tilde{\Gamma}_\mathrm{S}(N)$ for a monomer ($N = 1$) and a dimer ($N = 2$) above a silver surface (Sur) and in free space (FS). (c), (d) Decomposition of the transition rates into zero-phonon (ZPL) and phonon-assisted (PSB) components for the dimer $\tilde{\Gamma}^{\mathrm{ZPL/PSB}}_\mathrm{S}(2)$ and the monomer $\tilde{\Gamma}^{\mathrm{ZPL/PSB}}_\mathrm{S}(1)$. (e) Polariton spectral density $J_\mathrm{P}(\omega)$ for molecules above a silver surface and in free space. (f) $\eta$ dependence of $\mathcal{C}_{\mathrm{ZPL}}(\omega - \tilde{\omega}_\mathrm{S})$ (upper panel) and $\mathcal{C}_{\mathrm{PSB}}(\omega - \tilde{\omega}_\mathrm{S})$ (lower panel) of the FCWD for a dimer.}}
    
    \label{Fig:Coupling}
\end{figure*}

We focus on molecular aggregates that are weakly coupled to polaritons, and the interactions among the monomers are dominated by free-space dipole-dipole interactions. To derive the superradiance rate of a molecular aggregate, we first transform the total Hamiltonian in Eq.~(\ref{Eq:H_tot}) into the polaron frame, simplifying the treatment of exciton-phonon interactions, and derive the equation of motion from the quantum Liouville equation. Note that the superradiance rate of a molecular aggregate cannot be obtained by a standard second-ordered polaron quantum master equation due to the strong intermolecular dipole-dipole interactions. Instead, we transform the quantum Liouville equation into a modified rotating-frame. Then, we treat phonons and polaritons as baths, apply the Born-Markov and secular approximations \textcolor{black}{(weak coupling between molecules $\hat{\sigma}^-_\alpha$ and polaritons $\mathbf{\hat{f}}(\bf{r},\omega)$)}, derive a Lindblad-type quantum master equation, and finally identify the collective transition rate $\tilde{\Gamma}_{k}$ from the aggregate excited state $\ket{\tilde{\mathrm{E}}_{k}}$ to the ground state $\ket{\mathrm{G}}$. A detailed derivation and discussion can be found in Supplemental Material (SM) \cite{SM}, and the final result of $\tilde{\Gamma}_{k}$ is given by: 
\begin{align}
\nonumber
    \tilde{\Gamma}_{k} &= 2 \pi \sum_{\alpha,\beta} U^*_{\alpha k} U_{\beta k} \int_0^\infty \dd{\omega} \biggl\{ J^{\mathrm{Em}}_{\mathrm{P}, \alpha \beta}(\omega) \mathcal{C}_{\mathrm{V},\alpha\beta}(-\omega+\tilde{\omega}_k) \\
    & \quad + J^{\mathrm{Abs}}_{\mathrm{P}, \alpha \beta}(\omega) \mathcal{C}_{\mathrm{V},\alpha\beta}(\omega+\tilde{\omega}_k) \biggr\}.
\label{Eq:rate_kG}
\end{align} 
The modified polariton spectral densities are defined as $J^{\mathrm{Em}}_{\mathrm{P}, \alpha \beta}(\omega) = J_{\mathrm{P}, \alpha \beta}(\omega) \left[ n_{\mathrm{P}}(\omega) + 1 \right]$ and $J^{\mathrm{Abs}}_{\mathrm{P}, \alpha \beta}(\omega) = J_{\mathrm{P}, \alpha \beta}(\omega) n_{\mathrm{P}}(\omega)$, which correspond to the polariton emission and absorption processes, respectively. The thermal occupation number of polaritons is given by $n_{\mathrm{P}}(\omega) = \left[ e^{\hbar \omega / (k_\mathrm{B} T_\mathrm{P})} - 1 \right]^{-1}$, where $k_\mathrm{B}$ and $T_\mathrm{P}$ denote the Boltzmann constant and the temperature of the polariton bath, respectively. The Franck-Condon weighted density of states (FCWD) $\mathcal{C}_{\mathrm{V},\alpha\beta}(\omega) = \frac{1}{2 \pi} \int_{-\infty}^\infty \dd{t} e^{i\omega t} C_{\mathrm{V},\alpha\beta}(t)$ is the Fourier transform of the phonon time correlation function $C_{\mathrm{V},\alpha\beta}(t) = e^{-\phi_\mathrm{V}(0) + \phi_\mathrm{V}(t) \delta_{\alpha \beta}}$, where the phonon phase function is given by $\phi_\mathrm{V}(t) = \int_0^\infty \dd{\omega} \frac{J_{\mathrm{V}}(\omega)}{\omega^2} \left[ \coth{(\frac{\hbar \omega}{2 k_\mathrm{B} T_\mathrm{V}})} \cos{(\omega t)} - i\sin{(\omega t)} \right]$. Here, $T_\mathrm{V}$ denotes the temperature of all the independent phonon baths. The transition energy of aggregate state $k$ is expressed as $\hbar\tilde{\omega}_k = \hbar \omega_\mathrm{M} + e^{-\phi_\mathrm{V}(0)} \hbar \tilde{\Delta}^0_k
$, where $\hbar \tilde{\Delta}^0_k = \sum_{\alpha, \beta} U^*_{\alpha k} U_{\beta k} V^{0}_{\alpha \beta}$. 
 
The collective transition rate $\tilde{\Gamma}_{k}$ recovers the results of several previous studies under various limiting conditions, including (i) the single-monomer case \cite{Roy-Choudhury2015}, (ii) the absence of exciton-phonon coupling \cite{Chuang2024b}, and (iii) the free-space scenario without exciton-phonon coupling \cite{Spano1989}. This consistency supports the validity of our approach, and detailed demonstrations are provided in SM \cite{SM}. \textcolor{black}{We note that Eq. (\ref{Eq:rate_kG}) provides a robust description in the weak light-matter coupling regime, which is characterized by Markovian dynamics. Under strong light-matter coupling, the dynamics of the aggregate are instead expected to exhibit Rabi oscillations \cite{Wang2019, Chuang2024}.}

\textit{\textcolor{black}{System setup}}.--- 
We consider a minimal model composed of two monomers with an energy gap of $\hbar\omega_\mathrm{M} = 3.71$ eV and a transition dipole moment magnitude of $\abs{\boldsymbol{\mu}} = 10$ Debye, positioned above a silver surface in a head-to-tail (J aggregate) configuration, as depicted in Fig.~\ref{Fig1}(a). The monomers are separated by 1 nm, and the molecule-surface distance is set to 10 nm \textcolor{black}{to prevent the intrinsic loss mechanisms in the metal from dominating the decay process \cite{Ford1984}} and to ensure the system remains within the weak exciton-polariton coupling regime \cite{Chuang2024b}. \textcolor{black}{A more detailed analysis of the exciton-polariton coupling regime can be found in the previous work \cite{Wang2020}.} In this setup, the monomers are coupled to a global polariton bath, while each monomer is individually coupled to an independent phonon (vibrational) bath, as illustrated in Fig.~\ref{Fig1}(b). The dielectric function of silver is modeled using an analytical expression \cite{Melikyan2014} fitted to experimental data \cite{Johnson1972}. For the phonon spectral density, we employ a super-Ohmic form, i.e., $J_\mathrm{V}(\omega) = \eta \omega^3 \omega_{\mathrm{cut}}^{-2} e^{-\omega / \omega_{\mathrm{cut}}}$, where $\eta$ is a dimensionless \textcolor{black}{exciton-phonon} coupling constant \textcolor{black}{(not related to the exciton-polariton coupling strength)}, and $\omega_{\mathrm{cut}}$ denotes the cutoff frequency, fixed at $\omega_{\mathrm{cut}} = 100$ cm$^{-1}$ throughout this study. Note that our theoretical framework applies to arbitrary dielectric environments and is not restricted to the super-Ohmic form of the phonon spectral density. This setup is chosen to reveal an anomalous superradiance behavior. We focus on the transition rate of the superradiant state, i.e., $\tilde{\Gamma}_k = \tilde{\Gamma}_\mathrm{S}$, where the expansion coefficients satisfy $U_{\alpha \mathrm{S}} = 1/\sqrt{2}$ for all $\alpha$. This state corresponds to the lowest-energy excited state of a J aggregate. Moreover, we set the temperature $T_\mathrm{P} = T_\mathrm{V} = T$.

\textit{\textcolor{black}{Polaron-polariton effect on superradiance}}.---
\textcolor{black}{First, to explore the effect of polarons and polaritons on molecular superradiance, we vary the exciton-phonon coupling strength $\eta$ and place molecules above a silver surface to introduce exciton-polariton coupling. Fig.~\ref{Fig:Coupling}(a) presents the rate enhancement, defined as the transition rate of the dimer $\tilde{\Gamma}_\mathrm{S}(2)$ normalized by that of the monomer $\tilde{\Gamma}_\mathrm{S}(1)$, as a function of $\eta$ at $T = 0~\mathrm{K}$ for systems with exciton-polariton coupling (silver surface) and without it (free space).} In free space, the rate enhancement decreases monotonically as $\eta$ increases; in contrast, when above a silver surface, the enhancement initially increases before eventually decreasing. This initial rise in rate enhancement highlights a key distinction between the polariton-free and polariton-assisted superradiance, and appears to contradict the conventional understanding that exciton-phonon coupling generally suppresses superradiance. To investigate the origin of this anomalous behavior, we plot the transition rates of the dimer and monomer separately in Fig.~\ref{Fig:Coupling}(b). It is shown that the trend observed in the rate enhancement closely follows that of the dimer’s superradiance rate $\tilde{\Gamma}_\mathrm{S}(2)$. 

To better understand the influence of phonons on the superradiance rate, we decompose the FCWD in Eq.~(\ref{Eq:rate_kG}) into the zero-phonon line (ZPL) and the phonon sideband (PSB)  \cite{Jang2002, Iles-Smith2017, Clear2020, Pandey2024} as $\mathcal{C}_{\mathrm{V},\alpha\beta}(\omega) = \mathcal{C}_{\mathrm{ZPL}}(\omega) + \mathcal{C}_{\mathrm{PSB}}(\omega) \delta_{\alpha \beta}$, with
\begin{align}
    \mathcal{C}_{\mathrm{ZPL}}(\omega) &= \frac{1}{2 \pi} \int_{-\infty}^\infty \dd{t} e^{i\omega t} C_{\mathrm{V}, \alpha\alpha}(\infty), \\
    \mathcal{C}_{\mathrm{PSB}}(\omega) &= \frac{1}{2 \pi} \int_{-\infty}^\infty \dd{t} e^{i\omega t} \left[ C_{\mathrm{V}, \alpha\alpha}(t) - C_{\mathrm{V}, \alpha\alpha}(\infty) \right],
\end{align}
where $C_{\mathrm{V}, \alpha\alpha}(\infty) = e^{-\phi_\mathrm{V}(0)} = C_{\mathrm{V}, \alpha\beta{(\beta\neq\alpha)}}(t)$\textcolor{black}{, reflecting the absence of correlations between different phonon baths that each molecule couples to}. Note that the ZPL and PSB correspond to the zero-phonon and phonon-assisted transitions, respectively. For the treatment of $J_{\mathrm{P},\alpha\beta}(\omega)$, the translation symmetry of each monomer above the surface and the close proximity of the monomers allow us to define $J_{\mathrm{P}}(\omega) = J_{\mathrm{P},\alpha\alpha}(\omega) = J_{\mathrm{P},\beta\beta}(\omega) \approx J_{\mathrm{P},\alpha\beta}(\omega)$ \cite{Chuang2024b}. Therefore, at zero temperature, where $n_\mathrm{P}(\omega)=0$, the collective transition rate in Eq.~(\ref{Eq:rate_kG}) simplifies to $\tilde{\Gamma}_\mathrm{S}(N) \approx 2\pi \sum_{\alpha,\beta} U^*_{\alpha \mathrm{S}} U_{\beta \mathrm{S}} \int_0^\infty \dd{\omega} J_{\mathrm{P}}(\omega) \mathcal{C}_{\mathrm{ZPL}}(\omega-\tilde{\omega}_\mathrm{S}) + 2\pi \int_0^\infty \dd{\omega} J_{\mathrm{P}}(\omega) \mathcal{C}_{\mathrm{PSB}}(\omega-\tilde{\omega}_\mathrm{S})$. We further define the zero-phonon and phonon-assisted transition rates as $\tilde{\Gamma}^\mathrm{ZPL/PSB}_\mathrm{S}(N) =  2 \pi \int_0^\infty \dd{\omega} J_{\mathrm{P}}(\omega) \mathcal{C}_{\mathrm{ZPL/PSB}}(\omega-\tilde{\omega}_\mathrm{S})$, such that the total transition rate can be expressed as 
\begin{align}
    \tilde{\Gamma}_\mathrm{S}(N) \approx  \abs{\sum_{\alpha} U_{\alpha \mathrm{S}}}^2 \times \tilde{\Gamma}^\mathrm{ZPL}_\mathrm{S}(N) + \tilde{\Gamma}^\mathrm{PSB}_\mathrm{S}(N).   
\end{align} 
\textcolor{black}{The absence of a prefactor in the phonon-assisted transition rate arises from the fact that each molecule is coupled to its own independent local phonon bath.}
For a dimer, $\tilde{\Gamma}_\mathrm{S}(2) \approx 2 \tilde{\Gamma}^\mathrm{ZPL}_\mathrm{S}(2) + \tilde{\Gamma}^\mathrm{PSB}_\mathrm{S}(2)$. The overall behavior of $\tilde{\Gamma}_\mathrm{S}(2)$ in Fig.~\ref{Fig:Coupling}(b) thus reflects the combining effect of $\tilde{\Gamma}^\mathrm{ZPL}_\mathrm{S}(2)$ and $\tilde{\Gamma}^\mathrm{PSB}_\mathrm{S}(2)$, and we plot them separately in Fig.~\ref{Fig:Coupling}(c). 

\textcolor{black}{Figure~\ref{Fig:Coupling}(c) shows that,} in free space, increasing $\eta$ naturally enhances the phonon-assisted transition rate $\tilde{\Gamma}^\mathrm{PSB}_\mathrm{S}(2)$ \textcolor{black}{but suppresses} the zero-phonon transition rate $\tilde{\Gamma}^\mathrm{ZPL}_\mathrm{S}(2)$, and the relative changes in these two processes are nearly equal in magnitude. Since $\tilde{\Gamma}^\mathrm{ZPL}_\mathrm{S}(2)$ contributes twice in $\tilde{\Gamma}_\mathrm{S}(2)$, the overall superradiance rate is suppressed as $\eta$ increases. \textcolor{black}{On the other hand,} above the surface, $\tilde{\Gamma}^\mathrm{ZPL}_\mathrm{S}(2)$ and $\tilde{\Gamma}^\mathrm{PSB}_\mathrm{S}(2)$ both exhibit non-monotonic behavior. 
\textcolor{black}{Furthermore, by decomposing $\tilde{\Gamma}_\mathrm{S}(1) = \tilde{\Gamma}^\mathrm{ZPL}_\mathrm{S}(1) + \tilde{\Gamma}^\mathrm{PSB}_\mathrm{S}(1)$, Fig.~\ref{Fig:Coupling}(d) shows that the key distinction between the monomer and dimer responses lies in the monotonic decrease of $\tilde{\Gamma}^\mathrm{ZPL}_\mathrm{S}(1)$ with increasing $\eta$ when above the surface. This can be understood by the analytical solution $\tilde{\Gamma}^\mathrm{ZPL}_\mathrm{S}(1) = e^{-\eta} \times 2 \pi J_\mathrm{P}(\omega_\mathrm{M})$, which shows that its magnitude decays exponentially as $\eta$ increases.} 

To elucidate \textcolor{black}{the origin of the non-monotonic trend of $\tilde{\Gamma}^\mathrm{ZPL}_\mathrm{S}(2)$}, we examine the components of $\tilde{\Gamma}^\mathrm{ZPL}_\mathrm{S}(2)$, including $J_\mathrm{P}(\omega)$ and $\mathcal{C}_{\mathrm{ZPL}}(\omega-\tilde{\omega}_\mathrm{S})$, as shown in Figs.~\ref{Fig:Coupling}(e) and \ref{Fig:Coupling}(f). As $\eta$ increases, the peak of $\mathcal{C}_{\mathrm{ZPL}}(\omega - \tilde{\omega}_\mathrm{S})$ shifts to higher energy due to the reduction in the free-space dipole-dipole interaction $V^0_{12}$ between the monomers, i.e., $\hbar \tilde{\omega}_\mathrm{S} = \hbar \omega_\mathrm{M} + e^{-\eta} V^0_{12}$. This shift initially improves the resonance between $\mathcal{C}_{\mathrm{ZPL}}(\omega - \tilde{\omega}_\mathrm{S})$ and $J_\mathrm{P}(\omega)$ but eventually results in off-resonance. Although the magnitude of $\mathcal{C}_{\mathrm{ZPL}}(\omega - \tilde{\omega}_\mathrm{S})$ decreases monotonically with $\eta$, the resonance-induced enhancement dominates at small $\eta$, leading to the initial rise of $\tilde{\Gamma}^\mathrm{ZPL}_\mathrm{S}(2)$. The rate $\tilde{\Gamma}^\mathrm{PSB}_\mathrm{S}(2)$ in Fig.~\ref{Fig:Coupling}(c) exhibits a similar non-monotonic $\eta$ dependence with a slight difference that the enhancement persists over a broader range of $\eta$. This distinction is explained by the profile of $\mathcal{C}_{\mathrm{PSB}}(\omega - \tilde{\omega}_\mathrm{S}) 
$ \textcolor{black}{in Fig.~\ref{Fig:Coupling}(g): first,} $\mathcal{C}_{\mathrm{PSB}}(\omega - \tilde{\omega}_\mathrm{S})$ is broadband and thus overlaps a wider resonance window with $J_\mathrm{P}(\omega)$ than the sharply peaked $\mathcal{C}_{\mathrm{ZPL}}(\omega - \tilde{\omega}_\mathrm{S})$\textcolor{black}{; second,} the amplitude of $\mathcal{C}_{\mathrm{PSB}}(\omega - \tilde{\omega}_\mathrm{S})$ initially grows as $\eta$ rises from $0$ to $1.2$, while that of $\mathcal{C}_{\mathrm{ZPL}}(\omega - \tilde{\omega}_\mathrm{S})$ decreases monotonically.

The above results show that predicting the pattern of the polaron-polariton assisted superradiance is nontrivial. Nevertheless, we offer a transparent and physically intuitive way for analysis by evaluating: (i) the resonance condition between the FCWD and the polariton spectral density and (ii) the relative magnitude of the FCWD. \textcolor{black}{In addition, since the dipole-dipole interaction $V^0_{12} \propto \abs{\boldsymbol{\mu}}^2$, a larger transition dipole moment of the monomer typically leads to a more significant activation effect. Note that our theoretical framework can naturally accommodate disorder \cite{Spano2005, Chen2022, Catuto2025}, e.g., by varying monomer transition frequencies to model static disorder. The observed phenomenon is expected to persist as long as the energetic disorder remains small relative to the intermolecular dipole-dipole interaction. Conversely, comparable disorder and intermolecular coupling may lead to richer phenomena worthy of further investigation.}

\begin{figure}[!t]
    \centering
    \includegraphics[width=0.47\textwidth]{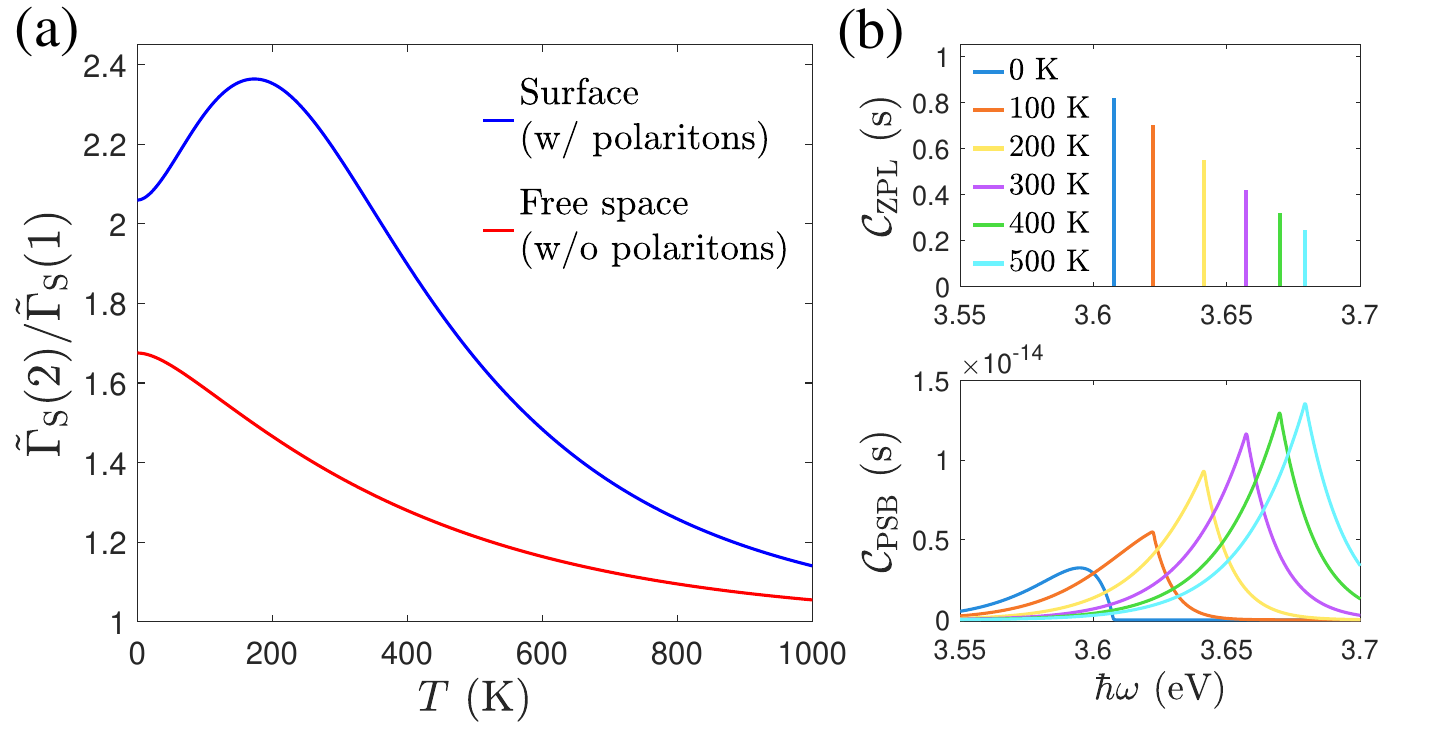} 
    \caption{(a) Temperature $T$ dependence of the superradiance rate enhancement $\tilde{\Gamma}_\mathrm{S}(2)/\tilde{\Gamma}_\mathrm{S}(1)$ for a dimer above a silver surface (with polaritons) and in free space (without polaritons), with $\eta = 0.2$. (b) $T$ dependence of $\mathcal{C}_{\mathrm{ZPL}}(\omega - \tilde{\omega}_\mathrm{S})$ (upper panel) and $\mathcal{C}_{\mathrm{PSB}}(\omega - \tilde{\omega}_\mathrm{S})$ (lower panel) of the FCWD for a dimer.}
    \label{Fig:T}
\end{figure}

\textit{\textcolor{black}{Temperature control of superradiance}}.---
While we have demonstrated the possibility of polaron-polariton assisted superradiance, directly controlling the exciton-phonon coupling strength in experiments remains challenging, hindering the verification of the predicted behavior. To overcome this difficulty, we propose an experimentally feasible alternative: tuning the temperature $T$, which can also effectively modulates the exciton-phonon coupling strength. Specifically, the phonon phase function can be rewritten as $\phi_\mathrm{V}(t) = \int_{-\infty}^\infty \dd{\omega} \frac{\eta'(\omega, T_\mathrm{V}) J_\mathrm{V}(\abs{\omega})}{\omega^2} e^{-i \omega t}$, where $\eta'(\omega, T) = \frac{1}{2} \left[ \coth{ ( \frac{\hbar \abs{\omega}}{2 k_\mathrm{B} T} )} + \mathrm{sign}(\omega) \right]$ serves as an effective $T$ and $\omega$ dependent exciton-phonon coupling strength function. Since $\coth{ ( \frac{\hbar \abs{\omega}}{2 k_\mathrm{B} T} )}$ increases monotonically with $T$, raising $T$ effectively enhances the exciton-phonon coupling strength. 

In Fig.~\ref{Fig:T}, we examine the effect of $T$ on superradiance with $\eta = 0.2$. The trend observed in the rate enhancement in Fig.~\ref{Fig:T}(a) is similar to that of the $\eta$-dependent case. The initial rise in the rate enhancement for a dimer above a silver surface can be understood by $\tilde{\Gamma}_\mathrm{S}(2) \approx 2\tilde{\Gamma}^{\mathrm{ZPL}}_\mathrm{S}(2) + \tilde{\Gamma}^{\mathrm{PSB}}_\mathrm{S}(2)$, where we have assumed $n_\mathrm{P}(\omega) \approx 0$ as $\hbar \tilde{\omega}_\mathrm{S} \gg k_\mathrm{B} T$. Both $\tilde{\Gamma}^{\mathrm{ZPL}}_\mathrm{S}(2)$ and $\tilde{\Gamma}^{\mathrm{PSB}}_\mathrm{S}(2)$ initially increase with $T$ due to the thermally induced blue shifts of $\mathcal{C}_{\mathrm{ZPL}}(\omega-\tilde{\omega}_\mathrm{S})$ and $\mathcal{C}_{\mathrm{PSB}}(\omega-\tilde{\omega}_\mathrm{S})$ [upper and lower panels of Fig.~\ref{Fig:T}(b)]. This spectral shift first brings FCWD into resonance with $J_\mathrm{P}(\omega)$ [Fig.~\ref{Fig:Coupling}(e)], enhancing the superradiance rate, but eventually drives it out of resonance, resulting in a turnover behavior.

The non-monotonic temperature dependence \textcolor{black}{presented in Fig.~\ref{Fig:T}} reveals an anomalous thermally activated superradiance in one-dimensional J aggregates.
Remarkably, this thermally induced enhancement becomes more pronounced and persists to higher temperatures, even up to room temperature, with increasing aggregate size \textit{N} (see SM \cite{SM}).
\textcolor{black}{In addition, the emission rate can exhibit superlinear or sublinear scaling with \textit{N}, or even decrease with increasing \textit{N}, depending on the alignment between the monomer transition energy and the surface plasmon polariton resonance \cite{Chuang2024b}.}
Moreover, the mechanism of this phenomenon differs fundamentally from that observed in earlier studies \cite{Renger2009, Eisfeld2017}. In our case, the behavior stems from the interplay between phonons and polaritons, whereas the previously reported effect originates solely from the monomer packing configuration. Notably, our theoretical framework is sufficiently general to capture both mechanisms.

\textit{Conclusion}.---
By combining macroscopic quantum electrodynamics and a modified polaron quantum master equation approach, we develop a general theory of superradiance in molecular excitons coupled to both phonons and polaritons at finite temperature. According to our theoretical approach, we demonstrate that superradiance can be enhanced by increasing either the exciton-phonon coupling strength or the temperature. This finding predicts an anomalous thermally activated superradiance in one-dimensional J aggregates, which is uncommon in free space and unexplained by previous theoretical frameworks. \textcolor{black}{Open questions remain, including quantifying the fraction of superradiant decay through radiative versus nonradiative channels in lossy media \cite{Liu2009, Zhao2020b} and maximizing the radiative fraction. \textcolor{black}{Furthermore, it is also intriguing to understand how polaritons influence superradiance under external driving fields and multiphoton excitation \cite{Raino2018, Liedl2024} through quantum master equations, as well as} in large aggregates where spatial inhomogeneity becomes significant \cite{Scully2006, Science2009}.} We believe that this work highlights the potential of dielectric environment engineering for tailoring collective light emission in molecular aggregates.

\begin{acknowledgments}
\textit{Acknowledgments}.---We thank Academia Sinica (AS-CDA-111-M02), National Science and Technology Council (111-2113-M-001-027-MY4) and Physics Division, National Center for Theoretical Sciences (112-2124-M-002-003) for the financial support.
\end{acknowledgments}

\bibliographystyle{apsrev4-2}
%

\end{document}